\documentclass[intlimits,sumlimits,twocolumn]{revtex4} 

\usepackage{dsfont}
\usepackage{amsmath}
\usepackage{graphicx}
\usepackage{color}

\DeclareMathOperator{\diag}{diag}		% Diagonale
\DeclareMathOperator{\trace}{tr}		% Spur
\newcommand{\e}[1]{\exp \left( #1 \right)}	% Exponentialfunktion

\begin{document}
\title{Credit Risk and the Instability of the Financial System: an Ensemble Approach}

\author{Thilo A. Schmitt}
\email{thilo.schmitt@uni-due.de}
\affiliation{Fakult\"at f\"ur Physik, Universit\"at Duisburg--Essen, Duisburg, Germany}
\author{Desislava Chetalova}
\affiliation{Fakult\"at f\"ur Physik, Universit\"at Duisburg--Essen, Duisburg, Germany}
\author{Rudi Sch\"afer}
\affiliation{Fakult\"at f\"ur Physik, Universit\"at Duisburg--Essen, Duisburg, Germany}
\author{Thomas Guhr}
\affiliation{Fakult\"at f\"ur Physik, Universit\"at Duisburg--Essen, Duisburg, Germany}

\date{\today}

\begin{abstract}
  The instability of the financial system as experienced in recent
  years and in previous periods is often linked to credit defaults,
  \textit{i.e.},  to the failure of obligors to make promised payments.
  Given the large number of credit contracts, this problem is
  amenable to be treated with approaches developed in statistical
  physics.  We introduce the idea of ensemble averaging and thereby
  uncover generic features of credit risk. We then show that the often
  advertised concept of diversification, \textit{i.e.},  reducing the
  risk by distributing it, is deeply flawed when it comes to credit risk. The risk of extreme losses
  remains due to the ever present correlations, implying a substantial and persistent
  intrinsic danger to the financial system.
\end{abstract}

\maketitle

\section{Introduction}

The past years demonstrated the devastating consequences
when financial markets collapse. The instability of the financial
system is closely connected to that of banks and related institutions
which, in turn, is directly coupled to the losses when the obligors,
\textit{i.e.}, the companies or individuals that borrowed money,
default and are unable to fully repay.  In the
recession following a market break down, a higher than usual number of
defaults occur~\cite{Crouhy2000}, severely worsening the situation. The
crisis of 2007--2009 was triggered by false assessment of the risk
involved with subprime mortgage credits \cite{Hull2009}.  The ensuing bankruptcy of
Lehman Brothers~\cite{Eichengreen2012} then released an avalanche
effecting the world economy as a whole.  Economists who saw the
problems piling up have pointed out the importance of improved credit
risk
estimation~\cite{bielecki2004credit,bluhm2003introduction,Duffie1999,lando2008credit,mcneil2005quantitative}.
However, a quantitative study satisfying the standards common in
physics is missing. Here, we want to close this gap by transferring a
standard tool from statistical physics, namely ensemble average, to
credit risk estimation.

\begin{figure}[htbp]
  \begin{center}
    \includegraphics[width=0.45\textwidth]{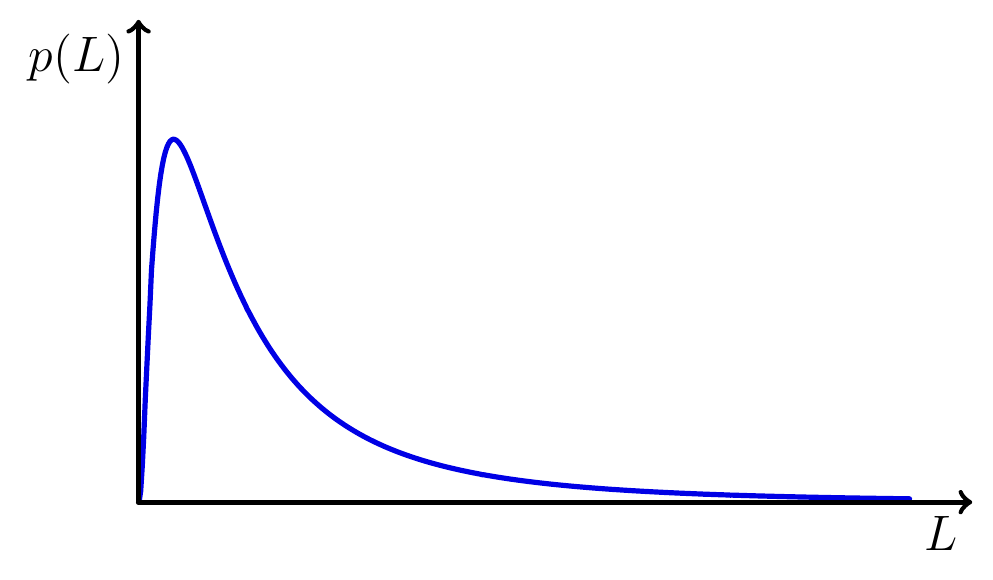}
  \end{center}
 \caption{Schematic drawing of a typical loss distribution $p(L)$
                 versus the relative loss $L$.}
 \label{draw}
\end{figure}

The problem can be traced back to the peculiar shape of a loss
distribution $p(L)$ for a portfolio consisting of a large number of
credit contracts. It is the probability density function (pdf) of the
dimensionless loss $L$ relative to the total exposure,
\textit{i.e.}, to the entire amount of money given out in the credits.
Typically, an empirical loss distribution looks as shown in
Fig.~\ref{draw}.
The asymmetry and the heavy tail on the right hand side are striking.
They are caused by the specific properties of the credit contracts:
The highest possible gain for the bank issuing the credits is only due
to interest and risk compensation, depending on the creditworthiness.
It occurs only if not a single credit defaults.  On the other hand, the
largest possible loss results from a complete loss of the lent money.
The danger lies in the heavy tail, which describes the probability for large losses that exceed the possible gains by far. Individual, large defaults such as
Enron or Lehmann Brothers as well as simultaneous defaults of many
small obligors as in the subprime mortgage crisis are the events
making this tail so heavy.

Thus, the issue of instability can be reformulated as the question
whether or not it is possible to get rid of this heavy tail. Financial
institutions often claim that this can be achieved by simply enlarging
the number of obligors and credit contracts in the portfolio. The
resulting \textit{diversification} is then believed to reduce the risk
for the bank. This view has been severely criticized, both with qualitative
reasoning \cite{Bhide} and quantitative studies addressing this important issue in the
economics literature, see \textit{e.g.}, \cite{Glasserman2004,schoenbucher2003credit,Heitfield2006,Glasserman2006}. 
Intuitively, it is not difficult to understand why the
concept of diversification is highly questionable. If the obligors are
correlated by some mutual dependencies, the default events will appear
clustered.  Only in the economically unrealistic case of zero
correlations, diversification can work.
Our goal is to identify generic features of credit risk using a standard approach from statistical physics: an ensemble approach for correlations.  As an
application, we then show that diversification is bound to fail. We
even derive an exact limiting loss distribution.

\section{Merton Model}
\begin{figure}[htbp]
  \begin{center}
    \includegraphics[width=0.45\textwidth]{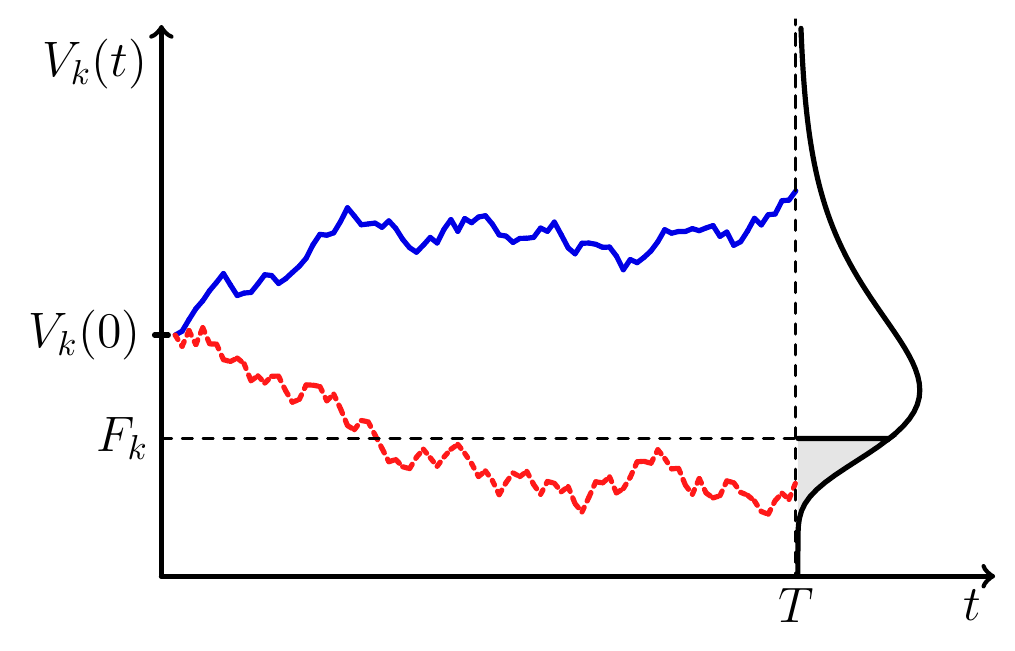}
  \end{center}
 \caption{Sketch of the Merton model. Default occurs if the economic state $V_k$ 
                 falls below the face value $F_k$ at maturity time $T$.}
 \label{fig:merton}
\end{figure}

We use a ``microscopic'' model, referred to as
``structural'' in economics, put forward by Merton~\cite{Merton1974}.
While ``reduced--form'' models, see \textit{e.g.},~\cite{Jarrow1997,Duffie1999,schoenbucher2003credit}, only provide an abstract description of default events,
the structural model traces defaults and losses back to stochastic processes describing the economic state of
each individual obligor. Suppose $K$ obligors hold credit contracts which bind them to pay back the
amount of money $F_k, \ k=1,\ldots,K$, referred to as face value, at the maturity time $T$.
The stochastic variable $V_k(t)$ is the economic state, \textit{i.e.}, the value of the $k$-th company, which can be retrieved after a bankruptcy and paid to the creditors.
As Fig.~\ref{fig:merton}
shows, the stochastic process now leads to a distribution of outcomes
$V_k(T)$ at maturity. In the cases $V_k(T)\ge F_k$, the obligor is
able to make the promised payment, the cases $V_k(T)< F_k$ are default
events.  This does not necessarily mean that the whole face value
$F_k$ is lost.  Rather, the normalized dimensionless loss of contract
$k$ is
\begin{align}
L_k = \frac{F_k - V_k(T)}{F_k}\Theta(F_k-V_k(T)) \ .
\label{indivloss}
\end{align}
The Heaviside function ensures that the loss is strictly positive,
because only the default events are to be taken into account.  The
entire credit portfolio comprises the individual losses of all $K$
contracts. The corresponding portfolio loss is the sum of the
individual losses $L_k$ weighted by their fraction $f_k$ in the
portfolio
\begin{align}
L = \sum_{k=1}^K f_k L_k \quad, \quad f_k = \frac{F_k}{ \sum_{i=1}^K F_i } \quad .
\label{eq:def}
\end{align}
The distribution of the portfolio loss is then given by
\begin{align}
p(L) & = \int d [V] g(V|\Sigma) \delta \left(L - \sum_{k=1}^{K} f_k L_k \right) \quad ,
\label{eq:loss}
\end{align}
where $g(V|\Sigma)$ is the multivariate distribution of the economic
states at maturity, $V=(V_1(T),\ldots,V_K(T))$. Importantly, it
depends on the $K\times K$ covariance matrix $\Sigma$ measuring the
above mentioned mutual dependencies between the obligors.  The
innocent--looking integral \eqref{eq:loss} is a highly non--trivial
object, first, because the $L_k$ contain, according to
Eq.~\eqref{indivloss}, a Heaviside function and, second, because the
multivariate distribution $g(V|\Sigma)$ of the economic states at
maturity is unknown.  It was Merton's seminal idea to estimate the
stochastic processes $V_k(t)$ by the stock prices $S_k(t)$, provided
all $K$ obligors are companies listed on the stock market. We assume
this from now on. In this Merton model it has been shown by numerical
simulations~\cite{Schafer2007,Koivusalo2011} that the heavy tail of
the loss distribution remains in the presence of even weak
correlations. In a less realistic setting, referred to as first
passage model with constant recovery and correlated defaults, this had
already been found in the economics
literature~\cite{Schonbucher2001,Glasserman2004}.

\section{Ensemble Average} 
\begin{figure}[htbp]
  \begin{center}
    \includegraphics[width=0.48\textwidth]{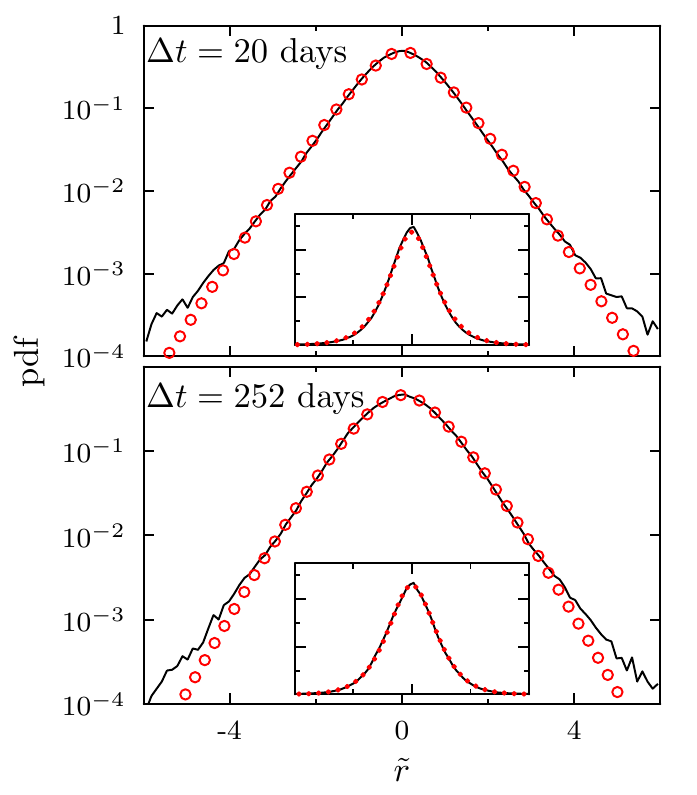}
  \end{center}
  \caption{Aggregated distribution of the rotated and rescaled returns
    $\tilde{r}$ for monthly returns $\Delta t = 20$ trading days (top)
    and yearly returns $\Delta t = 252$ trading days (bottom) for the
    model covariance matrix. The insets show the linear-linear plot.}
 \label{fig:pic4}
\end{figure}

Although the empirical stock market data
give us, in principle, access to $g(V|\Sigma)$, the formidable complexity
of the stock market dynamics keeps us still far from achieving a
generic understanding beyond the numerical case studies of
Ref.~\cite{Schafer2007}. The crucial problem is the non--stationarity
which is commonly studied for the returns, \textit{i.e.}, for the
dimensionless differences of the stock prices $S_k(t)$,
\begin{align}
r_k(t) = \frac{S_k(t+\Delta t)- S_k(t)}{S_k(t)} \ ,
\label{return}
\end{align}
where $\Delta t$ is referred to as return horizon.  Neither the
standard deviations or volatilities $\sigma_k$ of the individual
returns $r_k(t)$, nor the correlations $C_{kl}$ of any pair
$(r_k(t),r_l(t))$ are constant in time. We recall the relation
$\Sigma=\sigma C\sigma$ with $\sigma=\diag(\sigma_1,\ldots,\sigma_K)$
between covariance and correlation matrix.  To handle this
non--stationarity new concepts are called for. Here, we transfer the
idea of ensemble averaging from statistical physics.  To the best of
our knowledge, this whole approach is new in the economics literature.
According to Eq.~\eqref{eq:loss}, we need $g(V|\Sigma)$ which is
obtained from the more easily accessible distribution $g(r|\Sigma)$
depending on the $K$ component vector of the returns~\eqref{return}.
Recently, we constructed this distribution and confirmed its validity
by a careful data analysis~\cite{Chetalova2013}. We showed that
$g(r|\Sigma_s)$ is a \textit{multivariate} Gaussian,
\begin{align}
g(r|\Sigma_s) & = \frac{1}{ \sqrt{\det(2\pi\Sigma_s)}} \exp \left(-\frac{1}{2}r^\dagger\Sigma_s^{-1}r\right)  \ ,
\label{eq:multidist}
\end{align}
if the data are sampled in a data interval short enough that
$\Sigma_s$ is constant~\cite{Schmitt2013}. Importantly, we are interested in maturity
times $T$ of at least a month or so.  As the maturity time sets the
return horizon, $T=\Delta t$, the well--known heavy tails of the
\textit{individual} return distributions are not so pronounced yet.
Moreover, as we consider the \textit{multivariate} distribution of all
returns, the heavy tails of the \textit{individual} return
distributions are further suppressed.  However, we are interested not
in short, but in larger data intervals (not to be confused with the
return horizon) where sizable non--stationarity is present.  We take
it into account by averaging over the correlation matrices. We use the
Wishart distribution~\cite{Wishart1928} of the correlation matrices
$WW^\dagger$,
\begin{align}
w(W|C,N) & = \frac{\sqrt{N}^{KN}}
                                  {\sqrt{\det(2\pi C)}^N} \exp\left(-\frac{N}{2}\trace W^\dagger C^{-1}W\right) \ ,
\label{eq:wish}
\end{align}
which defines an ensemble of correlation matrices $WW^\dagger$ that
fluctuate around the mean correlation matrix $C$, calculated for the
entire data interval. The model matrices $W$ have dimension $K\times
N$, where $N$ formally corresponds to the length of the model time
series. It is a free parameter to be determined from the data later
on.  It governs the variance of the distribution~\eqref{eq:wish} and
thus the strength of the fluctuations around $C$. We break the
entire data interval into many short ones for which the
observation~\eqref{eq:multidist} is justified and we put
$\Sigma_s=\sigma WW^\dagger\sigma$. The average
\begin{align}
\langle  g \rangle (r|\Sigma,N) & = \int  d [W] w(W|C,N)  g (r|\sigma WW^\dagger\sigma)
\label{eq:ansatz}
\end{align}
then accounts for the non--stationarity. The result only depends on
$\Sigma=\sigma C\sigma$, calculated over the entire data interval, it
reads
\begin{align}
\langle  g \rangle(r|\Sigma,N) & = \frac{\sqrt{N}^K}
                                                                    {\sqrt{2}^{N-2}\Gamma(N/2)\sqrt{\det(2\pi\Sigma)}}\notag \\
& \frac{\mathcal{K}_{(K-N)/2} \left(\sqrt{Nr^\dagger\Sigma^{-1}r}\right)}
            {\sqrt{Nr^\dagger\Sigma^{-1}r}^{(K-N)/2}}
\label{eq:genresult}
\end{align}
with the Bessel function $\mathcal{K}$ of the second kind of order
$(K-N)/2$. We demonstrated the validity of
this result by obtaining $\Sigma$ directly from the data and
by fitting $N$~\cite{Schmitt2013,Chetalova2013}. Here, however, it is advantageous to make the
additional approximation that all off--diagonal correlation matrix
elements are equal, \textit{i.e.}, $C_{kl}=c, \ k\neq l$, hence
\begin{align}
C=(1-c) \mathds{1}_K + c ee^\dagger \ ,
\label{eq:C}
\end{align}
where $\mathds{1}_K$ is the $K\times K$ unit matrix and $e$ is a $K$
component vector with unity in all entries. By averaging all
off--diagonal matrix elements of $C$ measured in the whole data
interval, we find $c=0.26$ for monthly and $c=0.28$ for yearly returns. The data set consists of 306 stocks from the S\&P 500-index in the
time interval from 1992 to 2012~\cite{yahoo}. To test our result~\eqref{eq:genresult} with the
approximation~\eqref{eq:C}, we rotate the returns into the eigenbasis
of $\Sigma$ and scale with the eigenvalues. Integrating out all but one
rescaled return, denoted $\tilde{r}$, we have
\begin{align}
\langle g \rangle(\tilde{r}|N) & = \frac{\sqrt{ 2 }^{1-N}\sqrt{N}}
           {\sqrt{\pi} \, \Gamma(N/2)}\sqrt{N\tilde{r}^2}^{(N-1)/2} 
                       \mathcal{K}_{(N-1)/2} \left(\sqrt{N\tilde{r}^2}\right) \ .
\label{rk}
\end{align}
Figure~\ref{fig:pic4} shows the fit to the data which determines the value of $N$.
The fit for return horizons $\Delta
t$ of a month or a year --- which later on carries over to the
maturity times $T$ --- is very good with $N=4.2$ and $N=6.0$,
respectively. The tails are heavy, deviations only appear beyond the
third decade.

\section{Averaged Loss Distribution} The merit of the above
construction is the drastic reduction in the number of degrees of
freedom.  The return distribution of the highly complex,
non--stationary market is now fully and quantitatively characterized
by the two parameters $c$ and $N$ measuring the mean and the variance
of the fluctuations. This identification of generic features out of a
very large number of quantities is reminiscent of statistical
mechanics where a few macroscopic variables characterize a large
system that is microscopically described by a huge number of
variables. Hence, we are now able to also uncover generic features of
credit risk. Here, we greatly expand our previous work~\cite{Munnix2011a} to the more realistic case of a non-zero average correlation level and, in addition, we give an empirical verification of our approach. According to Eq.~\eqref{eq:loss}, the averaged loss
distribution reads
\begin{align}
\langle p\rangle(L) & = \int d [V] \langle g\rangle (V|\Sigma,N) 
                        \delta \left(L - \sum_{k=1}^{K} f_k L_k \right) \ .
\label{avloss}
\end{align}
We infer the price distribution $\langle g\rangle (V|\Sigma,N)$ from
our result~\eqref{eq:genresult} for the returns. The assumption that
the stock prices $S_k(t)$ follow a Geometric Brownian Motion
with drift and volatility constants $\mu_k$ and $\rho_k$,
respectively, leads to a multivariate Gaussian of the
form~\eqref{eq:multidist} for the returns. Hence, this is fully 
consistent with the above ensemble approach. Our data comparison
thus strongly corroborates the usage of the Geometric Brownian Motion.

To make analytical progress, we write the return
distribution~\eqref{eq:genresult} as a Fourier integral in the $K$
component vector $\omega$
\begin{align}
\langle  g \rangle (r|\Sigma,N) &= \frac{1}{2^{N/2}\Gamma(N/2)} 
                                                 \int\limits_0^\infty dz \ z^{N/2-1} \exp\left(-\frac{z}{2}\right)
                                                                                       \notag\\
& \quad \int \frac{d[\omega]}{(2 \pi)^K} 
                   \exp \left(- \text{i} \omega \cdot r - \frac{z}{2N} \omega^\dagger\Sigma\omega\right)
\label{eq:pavg} \ .
\end{align}
We insert the approximation~\eqref{eq:C} and linearize
the square of the scalar product $\omega \cdot e$ in the exponent by
another Fourier transform in $u$, say. The $\omega$ integral is then
trivial and we find after straightforward steps
\begin{align}
  \langle g \rangle(r|c,N) & =  \frac{1}{ 2^{N/2} \Gamma( N/2 ) }  
    \frac{ 1 }{ \det \sigma }  \int_0^\infty dz \ z^{ N/2 - 1 } e^{-z/2} \notag \\
  & \ \times \sqrt{ \frac{ N }{ 2 \pi z } } \sqrt{ \frac{ N }{ 2 \pi z ( 1 - c) } }^K  
  \int_{-\infty}^{\infty} du \, \e{- \frac{ N  }{ 2 z } u^2 } \notag \\
  & \ \times \e{- \sum_{k=1}^K \frac{ N }{ 2 z ( 1 - c)
      \sigma_k^2 } ( r_k + \sqrt{ c } u \sigma_k )^2 } \, .
\label{eq:avg_result}
\end{align}
Following Merton but extending his idea to our ensemble
approach, we now use the averaged return distribution to estimate the
averaged distribution $\langle g\rangle (V|\Sigma,N)$ of the economic
states $V_k(T)$ at maturity. According to It\=o's Lemma~\cite{Ito1944} we set
\begin{align}
r_k \longrightarrow \ln\frac{V_k(T)}{V_{k0}} - \left(\mu_k-\frac{\rho_k^2}{2}\right)T 
\label{log}
\end{align}
with $V_{k0}=V_k(0)>0$. We notice $\sigma_k=\rho_k\sqrt{T}$. Inserting this into
Eq.~\eqref{avloss} yields the exact expression for the averaged loss
distribution within our model.

To enforce simplicity, we consider a credit portfolio in which
all face values are of the same order, implying that $f_k \approx
1/K$. As the number of obligors $K$ is large, we may safely carry
out a second order approximation in $1/K$. Again, after tedious but
straightforward steps, we eventually arrive at the double integral
representation
\begin{align}
  \langle p \rangle(L|c,N) & = \frac{ 1 }{ \sqrt{ 2 \pi } 2^{N/2} \Gamma( N/2 ) }   \int_0^\infty dz \ z^{N/2 -1} e^{-z/2}   \sqrt{ \frac{ N }{ 2 \pi } }\notag \\
  & \quad \times \int_{-\infty}^{+\infty} du \e{-\frac{ N }{ 2 } u^2 } \frac{ 1 }{ \sqrt{ M_2(z,u) } } \notag\\
  & \quad \times \e{ - \frac{ ( L - M_1(z,u) )^2 }{ 2 M_2(z,u) } }
\label{avlossapprox}
\end{align}
for the average loss distribution. Here, we have 
\begin{align}
M_1(z,u) & =  \sum_{k=1}^K f_k m_{1k}(z,u)  \notag\\
M_2(z,u) & =  \sum_{k=1}^K f_k^2 \left(m_{2k}(z,u) - m_{1k}(z,u)^2\right) 
\label{largems}
\end{align}
with the moments of order $j=1,2$ given by
\begin{align}
  m_{jk}(z,u) & = \frac{ \sqrt{N} }{ \rho_k \sqrt{ 2 \pi T ( 1 - c ) } }  \int_{-\infty}^{\hat{F}_k} d \hat{V}_k \  \notag \\
  & \left( 1 - \frac{V_{k0} }{ F_k } \e{ \sqrt{z} \hat{V}_k +  \left( \mu_k - \frac{ \rho_k^2 }{ 2 } \right)T }  \right)^j \notag \\
  & \qquad \times \e{- \frac{ \left( \hat{V}_k + \sqrt{cT} u \rho_k
      \right)^2 }{ 2 T ( 1 - c) \rho_k^2 / N } } \ .
\label{moments} 
\end{align}
The upper bound of integration is $\hat{F}_k = (\ln(F_k/V_{k0}) - (
\mu_k - \rho_k^2/2)T)/\sqrt{z}$ and the change of variables leads to $
\hat{V}_k=(\ln(V_k(T)/V_{k0}) - (\mu_k-\rho_k^2/2)T )/\sqrt{z}$.
 Since the integral can be expressed
in terms of special functions, we are left with only the $z$ and $u$
integrals which have to be evaluated numerically.
\begin{figure}[ht!]
  \begin{center}
    \includegraphics[width=0.45\textwidth]{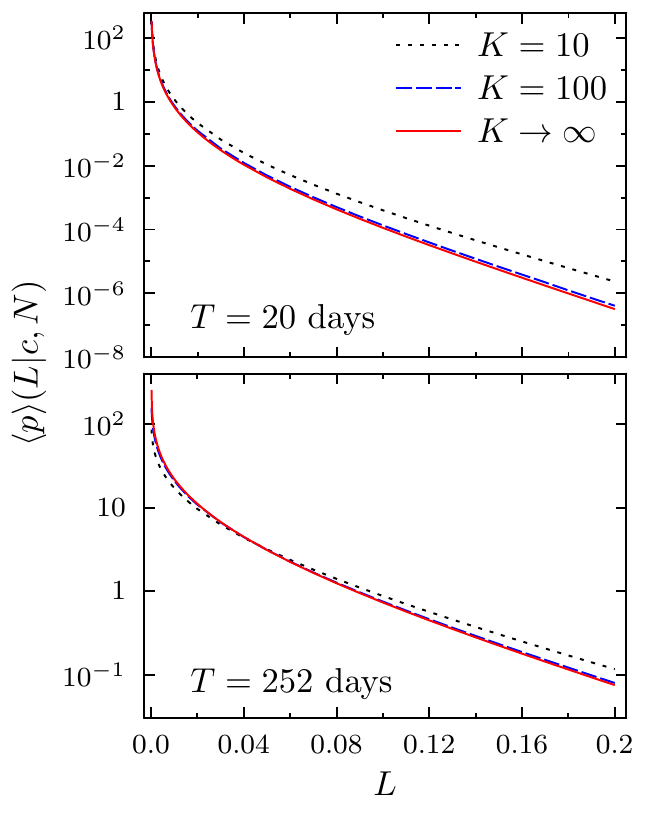}
  \end{center}
  \caption{The average loss distribution for $N=4.2$ and a correlation
    level of $c=0.26$ with drift $\mu_0=0.013$ $\text{month}^{-1}$,
    volatility $\rho_0=0.1$ $\text{month}^{-1/2}$, $T=20$ trading days (top) and 
    for $N=6$, a correlation
    level of $c=0.28$ with drift $\mu_0=0.17$ $\text{year}^{-1}$,
    volatility $\rho_0=0.35$ $\text{year}^{-1/2}$ and $T=1$ year (bottom). The
    dotted and dashed line are for $K=10$ and $K=100$, while the solid
    line shows the limit $K\rightarrow \infty$.}
 \label{fig:kdep}
\end{figure}

Further simplifications occur in the interesting case of a homogeneous
credit portfolio, in which all parameters for the obligors are the
same, face value $F_k=F_0$, variance $\rho^2_k=\rho_0^2$, drift
$\mu_k=\mu_0$ and initial value $V_{k0}=V_0$. Of course, this does not
mean that the actual paths explored by the stochastic processes are
the same.  The fractions are now all equal, $f_k = 1/K$, the same is
true for the moments $m_{jk}(z,u)=m_{j0}(z,u)$.  Importantly, this allows us to calculate the strict limit $K\to\infty$ of the averaged
loss distribution~\eqref{avloss}. The Gaussian in
Eq.~\eqref{avlossapprox} becomes a $\delta$ function such that
\begin{align} 
  \lim_{K\to\infty} \langle p \rangle(L|c,N) & = \frac{ 1 }{ 2^{N/2} \Gamma( N/2)
  } \sqrt{ \frac{ N }{ 2 \pi } } \int_0^\infty dz \, z^{N/2-1}
  e^{-z/2}\notag \\
  & \ \times  \int_{-\infty}^{+\infty} du \exp\left( - \frac{ N }{ 2 } u^2 \right) \delta ( L - m_{10}(z,u) ) \notag\\
  & =  \frac{ 1 }{ 2^{N/2} \Gamma( N/2) } \sqrt{ \frac{ N }{ 2 \pi } } \int_0^\infty dz \, z^{N/2-1} e^{-z/2} \notag\\
  & \ \times \exp\left( - \frac{ N }{ 2 } u_0^2 \right) \frac{ 1
  }{ \left| \partial m_{10}(z,u) / \partial u \right|_{z,u_0} } \ ,
\label{limres}
\end{align}
where $u_0$ is the zero of the first moment, $m_{10}(z,u_0)=0$. This
partly implicit formula yields a strict lower bound for the tail of
the averaged loss distribution.

\section{Results} In Fig.~\ref{fig:kdep} we show our
result~\eqref{avlossapprox} for homogeneous credit portfolios of
different sizes $K=10,100$, compared with the limiting curve
\eqref{limres} for a maturity time of $T=1$ month (top) and $T=1$ year (bottom). 
The parameters are chosen according to the data analysis, so we have $N=4.2$ and $N=6.0$. 
From the data set we find a drift of $\mu_0=0.013$ month$^{-1}$ and $\mu_0=0.17$ year$^{-1}$, volatility of $\rho_0=0.1$ month$^{-1/2}$ and $\rho_0=0.35$ year$^{-1/2}$ and an average correlation level of $c=0.26$ and $c=0.28$ depending on the time horizon. The face value is $F_0=75$ and the initial value is $V_0=100$. Both have the dimension currency. Notice that for the maturity time of one month the probability of a default is much smaller due to the lower volatility and the reduced time horizon.
One clearly sees that the tails of the
averaged loss distribution for finite $K$ very quickly reach the
limiting curve for $K\to\infty$. We thus arrive at the truly disturbing
observation that diversification  generically cannot work for any
realistic choice of correlation structure.

\section{Conclusions} We uncovered and derived generic features of the
loss distribution for credit portfolios. Our starting point was the
Merton model which is known to give a realistic description. By
transferring the concept of ensemble averaging to
this problem, we were able to derive an averaged portfolio loss distribution
which depends on only two parameters that fully account for the
non--stationary dynamics of the markets. Data analysis strongly
corroborates our approach. As an important application, we showed that
diversification is bound to fail for a homogeneous portfolio.  This is
due to the correlations which are always present in reality.
Pictorially speaking, they glue together the obligors and thereby let
them act to some extent like just one obligor. Thus, we have no reason
to hope that diversification can work for any other non--homogeneous
but realistic credit portfolio. All this is tantamount to saying that
there is an intrinsic instability in the markets which cannot be
overcome.  We emphasize that we did not study ``credit contagion'' or
avalanche effects after the onset of a crisis, as for example in
Ref.~\cite{Kuhn2003}. We uncovered the substantial, unavoidable
stability risk which is always there, even in periods in which the
markets appear quiet. It thus seems reasonable to advertize a
considerable enlargement of the equity held by the banks and
other creditors.

\end{document}